\documentclass{article}
\usepackage{pgfplots}
\usepackage{lipsum}
\usepackage{caption}
\usepackage{subcaption}
\usepackage{authblk}
\usepackage[english]{babel}
\usepackage{amsfonts}
\usepackage{amsmath}
\usepackage[top=2cm, bottom=2cm, left=2cm, right=2cm]{geometry}
\usepackage{fancyhdr}
\usepackage{multicol}
\usepackage{bbold}
\usepackage{epsfig}
\usepackage{graphics}
\usepackage{epsfig}
\usepackage{lipsum}
\usepackage{fancyhdr}
\usepackage{booktabs}
\usepackage{wrapfig}
\usepackage[export]{adjustbox}
\usepackage{rotating}
\usepackage{amsmath}

\usepackage{hyperref}
\usepackage{cleveref}

\usepackage{cite}
\renewenvironment{abstract}{\hfill\begin{minipage}{0.95\textwidth}
		\rule{\textwidth}{1pt}}
	{\par\noindent\rule{\textwidth}{1pt}\end{minipage}}

\makeatletter
\usepackage{orcidlink}
\makeatother

\begin{document}
	
	\title{\textbf{Multiple phase estimation with photon-added multi-mode coherent states of GHZ-type}}
	\author[1]{\textbf{H. Saidi}}
	\author[1,2]{\textbf{A. Slaoui \normalsize\orcidlink{0000-0002-5284-3240}}{\footnote {Corresponding author: {\sf abdallah.slaoui@um5s.net.ma}}}}
	\author[1,2]{\textbf{H. EL Hadfi}}
	
	\author[1,2]{\textbf{ R. Ahl Laamara \normalsize\orcidlink{0000-0002-8254-9085}}}
	\affil[1]{\small LPHE-Modeling and Simulation, Faculty of Sciences, Mohammed V University in Rabat, Rabat, Morocco.}
	\affil[2]{\small  Centre of Physics and Mathematics, CPM, Faculty of Sciences, Mohammed V University in Rabat, Rabat, Morocco.}

	\maketitle
	
\begin{abstract}
This paper explores multiparameter quantum metrology using Greenberger-Horne-Zeilinger (GHZ)-type photon-added coherent states (PACS) and investigates both independent and simultaneous parameter estimation with linear and non-linear protocols, highlighting the significant potential of quantum resources to enhance precision in multiparameter scenarios. To provide a comprehensive analysis, we explicitly derive analytical expressions for the quantum Cramér-Rao bound (QCRB) for each protocol. Additionally, we compare the two estimation strategies, examining the behavior of their QCRBs and offering insights into the advantages and limitations of these quantum states in various contexts. Our results show that simultaneous estimation generally outperforms independent estimation, particularly in non-linear protocols. Furthermore, we analyze how the QCRB varies with the coherent state amplitude \(|\alpha|^2\), the number of estimated parameters \(d\), and the photon excitation order \(n\) across three protocols. The results indicate that increasing \(|\alpha|^2\) and decreasing \(d\) improves estimation precision. For low \(n\), the variation in the QCRB is similar for both symmetric and antisymmetric cases; however, at higher \(n\), the antisymmetric case exhibits slightly better precision. The dependence on \(d\) is comparable for both types of states. We also compare PACS-based GHZ states with NOON states and entangled coherent states, demonstrating the relative performance of each. Finally, we conclude with an analysis of homodyne detection in the context of a linear protocol, discussing its impact on estimation accuracy.

\end{abstract}

\vspace{0.5cm}
\section{Introduction}
High-precision measurement plays a crucial role in any advanced quantum technologies; therefore, quantum metrology leverages quantum resources to enhance the precision of parameter estimation in physical systems \cite{Helstrom1969,2,3,4,5,6,10,11,12,13,14,15,16}. Its applications span various fields, from gravitational wave detectors \cite{19,20,21,22} to time-frequency standards \cite{23}, optical interferometry \cite{18,Demkowicz2015} and atomic spectroscopy \cite{Wineland1992}. Frequently, the measurement goal is to estimate a parameter of the system's Hamiltonian, and the limitations on our capacity for precise measurements are dictated by the Heisenberg uncertainty principle. In quantum estimation theory, the typical procedure involves preparing $N$ particles in an initial state (featuring entanglement or squeezing) which evolves over time before detection \cite{2,Holevo1982}. This process is iterated to collect statistics and derive an estimate of the uncertain parameter. The estimation error relies on the resources assigned to the probe and the protocol, specifically the number of particles ($N$), the duration of each repetition ($t$), and the number of repetitions ($\mu= T/t$ where $T$ is the total duration of the protocol). The mathematical framework used to assess the estimation error is primarily based on the quantum Cramér-Rao bound (QCRB); ${\rm Var}(\theta)\geq {\cal F}_{\theta}^{-1}$, which is dependent on the quantum Fisher information ${\cal F}_{\theta}$ (QFI) \cite{Braunstein1996,Slaoui2018,Dakir2023}. In the context of classical resources (uncorrelated product states), the central limit theorem indicates that the estimation error of a parameter $\theta$ scales according to the shot-noise limit (or standard limit), with ${\rm Var}\left(\theta\right)\sim1/\sqrt{N\mu}$ for large $N$. Additionally, optimal precision in estimating parameters is attained through a minor variance. Consequently, the primary objective of quantum estimation processes is to minimize the variance, aligning with the maximum QFI. To this end, surpassing the QCR bound is a real major challenge for quantum estimation theory\cite{Gerry2001,Bera2014,Choi2017,Modi2011,Sone2019,Xiang2013}. When using $N$ noninteracting particles prepared in a GHZ state, maximizing the variance of the Hamiltonian, it has been proven that the error surpasses the standard limit and scales as $1/(N\sqrt{\mu})$ with a customary $\sqrt{N}$ enhancement, known as the Heisenberg limit \cite{11,Toth2014,Szczykulska2016}.\par

As previously discussed, the fundamental cornerstone of quantum estimation theory lies in the QCRB, which consistently reaches saturation when only one parameter is being estimated. However, saturating this bound becomes intricate when multiple parameters are estimated simultaneously, leading to the replacement of ${\rm Var}\left(\theta\right)$ by the covariance matrix ${\rm Cov}(\hat{\theta})$ and the QFI by the QFI matrix $F_{\theta}$ (i.e., ${\rm Cov}(\hat{\theta})\geq F_{\theta}^{-1}$) \cite{18,Szczykulska2016, Ciampini2016,Proctor2018,Yousefjani2017,Ragy2016,Goldberg2021,Nichols2018,Liu2020,Liu2015}. This challenge arises from the incompatibility between the measurements of the various estimated parameters. Consequently, multiparametric quantum estimation has garnered significant interest and evolved into a critical undertaking across various contexts \cite{Albarelli2022,Ikken2023,Liu2017}. Unlike classical metrology, which concentrates on measuring independent parameters, multiparametric quantum metrology considers the influence of quantum correlations among the measured parameters. Recent advancements have enhanced our understanding of the saturation of QCR bound for pure states \cite{Fujiwara2001,Matsumoto2002}. Various multiparametric estimation protocols have been introduced in diverse scenarios, paving the way for novel prospects in multiparametric quantum metrology. Additionally, when multiple parameters are simultaneously estimated in a manner that surpasses individual estimation strategies, simultaneous estimation can yield greater precision compared to individual estimations \cite{Abouelkhir22023,Ragy2016}. Simultaneous and independent estimation in quantum metrology refers to the ability to measure several physical quantities simultaneously and independently with high precision \cite{Humphreys2013,Yue2014,Pinel,Yao2014,Yao,Lang}. In simultaneous estimation, the objective is to measure multiple physical quantities concurrently using an appropriate quantum state. This approach offers distinct advantages when the measured parameters are correlated or interact, potentially leading to enhanced precision compared to individual parameter estimation. In contrast, independent estimation focuses on measuring each physical quantity separately and independently. In this case, quantum states and measurement protocols adapted to each quantity are employed. Measurements are performed individually for each quantity, without interacting or interfering with other measured quantities. This protocol enables obtaining accurate information about each physical quantity in isolation.\par

The use of entangled quantum states in multiparametric quantum metrology offers significant advantages \cite{Sanders2012}. It enables more accurate measurements to be obtained by exploiting the quantum properties of these states, particularly their increased sensitivity to variations in the measured parameters. For instance, in a study conducted by Humphreys et al.\cite{Humphreys2013}, a generalized form of NOON states was utilized as a resource to approach the Heisenberg limit. The research demonstrated that simultaneous estimation using generalized NOON states outperformed independent estimation using the same type of states. However, it is essential to note that the NOON state is not the only state capable of achieving the Heisenberg limit. Another valuable state in this context is the entangled coherent state (ECS), which has been extensively applied and studied in the field of quantum metrology \cite{Liu2016,Joo2012,Jing2014,Joo2011}. Significantly, experiments conducted in Mach-Zehnder interferometers have demonstrated that entangled coherent states exhibit superior performance compared to NOON-type states \cite{Jing2014}. On the other hand, GHZ-states, which are  maximally entangled and possess extremely non-classical properties, are among the candidates that can achieve the Heisenberg limit \cite{Zukowsk1998,DHondt2004}. The GHZ state for a system of $n$-qubits may be expressed as $|GHZ\rangle_{n}=\left(|0\rangle^{\otimes{n}}+|1\rangle^{\otimes{n}}\right)/\sqrt{2}$. The $3$-qubit GHZ state is the most basic and already demonstrates non-trivial multipartite entanglement, described as
$|GHZ\rangle=\left(|000\rangle+|111\rangle\right)/\sqrt{2}$. It should be noted that generating GHZ states can be a technical challenge due to the need to preserve quantum entanglement during the creation and manipulation of the states. However, with recent advancements in quantum technologies, it has become possible to generate GHZ states with a high degree of precision and control \cite{Gerry1996,Pan2000}.\par

In quantum optics, photon states can be utilized to generate GHZ states. For instance, GHZ states can be created using single-photon sources and interferometers. In atomic physics, trapped atoms or ions can also be employed for GHZ state generation. In fact, by leveraging the interaction properties of atoms or ions, we can implement quantum logic gates and entwine qubits to achieve GHZ states. On the other hand, quantum information processing predominantly initially concentrated on discrete entangled states (i.e., the finite-dimensional states), such as the discrete levels of an atom or the polarizations of a photon. However, the transition from discrete to continuous variables has also demonstrated advantages in efficiently coding and manipulating quantum information. In this context, it is anticipated that coherent states, representing states with continuous variables, will play a pivotal role. Their appeal lies in their mathematical elegance, characterized by properties of continuity and over-completion, as well as their proximity to classical physical states, minimizing Heisenberg's uncertainty relation. In Ref.\cite{Gerry2007} the authors introduce a method for producing entangled coherent states of a two-mode field. Additionally, researchers have explored optimal quantum information processing through GHZ-type states. Jeong et al.\cite{Jeong2006} demonstrate the creation of GHZ-type entangled coherent states using beam splitters with single-mode coherent-state superpositions, showcasing violations of Bell inequalities for such states. Moreover, the concept of the photon-added coherent state was initially introduced in \cite{Agarwal1991}. The authors introduced a hybrid non-classical state that exhibited properties halfway between those of a strictly quantum Fock state and a classical coherent state. In a more recent development, to create a coherent state with a single photon added experimentally, Zavatta et al.\cite{Zavatta2004} used a single-photon detector with a balanced homodyne detector. This innovative approach enabled them to visually observe the transition process from classical to quantum states \cite{Zavatta2004}.\par

In this work, our emphasis is on optical interferometer schemes, both linear and nonlinear parameterization strategies, employing a generalized form of GHZ-type photon-added coherent states. We investigate their influence on quantum metrology and compare the performance of multiple phase estimations in individual and simultaneous metrological strategies within linear and non-linear protocols. This paper is structured as follows. In Sec.(2), we introduce multiple-mode GHZ photon-added coherent states. In Sec.(3), we examine the Cramer-Rao limit for independent and simultaneous multiparameter estimates. Next, in Sec.(4), we discussed the performance of these states and compared the results obtained with the entangled coherent (ECS) states and NOON states. Finally, We conclude this paper with the concept of homodyne detection in the case of a linear protocol. By
assuming that all parameters under estimation are very small, we derive the analytical expression
of the total variance according to the error propagation equation. 

\section{Multiparametric process in quantum metrology}
\subsection{Cramér-Rao bound and quantum Fisher information matrix}
One of the main issues in quantum metrology is estimating the small phase for the unit dynamics of a linear interferometer,  where $\rho_{\varphi}=e^{-i{\cal H}\varphi}\rho e^{i{\cal H}\varphi}$. 
The unitary transformation depending only on a single-phase parameter $\varphi$ causes the initial state $\rho$ to evolve into $\rho_{\varphi}$. Then a measurement is made, followed by data analysis to determine the small value of $\varphi$. 
The quantum Fisher information denoted as \({\cal F}_{Q}\left(\rho_{\varphi}\right)\), serves as a quantitative measure of how sensitive a quantum state is to changes in a parameter. It defines the ultimate limit of measurement precision. Specifically, the quantum Cramér-Rao bound states that the variance of any unbiased estimator of the parameter, $\left(\Delta \varphi\right)^{2}$, is restricted by the inverse of the Quantum Fisher Information, as described below.
\begin{equation}
\left(\Delta \varphi\right)^{2}\geq \frac{1}{\eta{\cal F}_{Q}\left(\rho_{\varphi}\right)},
\label{1}
\end{equation}
where $\eta$ stands for the number of independent measurements.\par

In a quantum multiparameter metrology process, the quantum state $\rho$ is contingent upon a collection of deterministic parameters $ \varphi= {\varphi_p}$. The estimation of these parameter values $\varphi$ is accomplished through the acquisition of observations during the measurement of the quantum system.  Importantly, for a given set of parameters \( \{\varphi_p\} \), the quantum Cramér-Rao inequality is formulated as follows \cite{Helstrom1969,3,Liu2016,Jing2014}:
\begin{equation}
C \geq F^{-1},
\label{2}
\end{equation}
F represents the Fisher information matrix, while C denotes the covariance matrix. The elements of these matrices are defined as follows
\begin{equation}
C_{pq} := \text{cov}(\hat{\varphi}_p, \hat{\varphi}_q) = \frac{1}{2}\left\langle\{\hat{\varphi}_p, \hat{\varphi}_q\}\right\rangle - \left\langle\hat{\varphi}_p\right\rangle\left\langle\hat{\varphi}_q\right\rangle,
\label{3}
\end{equation}
where ${\hat{\varphi}_p}$ represents the set of unbiased estimators, the symbol ${\{\cdot, \cdot\}}$ denotes the anticommutation. Let $L_p$ be the symmetric logarithmic derivative (SLD) for ${\varphi}_p$, which is a Hermitian operator satisfying
\begin{equation}
\frac{\partial \rho}{\partial \varphi_p} = \frac{1}{2}(\rho L_p + L_p \rho).
\label{4}
\end{equation}
Next, using the symmetric logarithmic derivative, the Fisher information matrix is defined as follows
\begin{equation}
F_{pq} = \frac{1}{2}\text{Tr}[(L_p L_q + L_q L_p)\rho].
\label{5}
\end{equation}
For a pure state \(\rho = |\psi\rangle\langle\psi|\), the elements of the Fisher information matrix are expressed as
\begin{equation}
F_{pq} = 4\text{Re}(\langle\partial_p\psi|\partial_q\psi\rangle - \langle\partial_p\psi|\psi\rangle\langle\psi|\partial_q\psi\rangle).
\label{6}
\end{equation}
Here, we employ the total variance, defined as \(|\delta\hat{\varphi}|^2 = \text{Tr}(C)\), a crucial quantity in the multiparameter estimation process. In this regard, applying the trace operator on the two members of inequality \eqref{2} yields to
\begin{equation}
|\delta\hat{\varphi}|^2 \geq \text{Tr}(F^{-1}).
\label{7}
\end{equation}
For a two-parameter scenario, equation \eqref{7} simplifies to
\begin{equation}
|\delta\hat{\varphi}|^2 \geq \frac{1}{F_{effective}}
\label{8}
\end{equation}
where \(F_{effective} = \text{det}(F)/\text{Tr}(F)\) is regarded as the effective quantum Fisher information. The symbol \(\text{det}(\cdot)\) denotes the determinant of a matrix.

\subsection{Parametrization process}
In a parametrization procedure involving p parameters, each parameter is represented by the Hermitian operator\cite{Liu2015,Liu2016}
\begin{equation}
\mathcal{H}_p = i(\partial_p U^\dagger_\varphi)U_\varphi.
\label{9}
\end{equation}
There are two basic techniques for detecting numerous parameters in a multiparameter unitary parametrization process: sequential protocols and simultaneous protocols. Consider a collection of Hermitian operators ${H_p}$. The description of the simultaneous protocol is provided by
\begin{equation}
U = \exp\left(\sum_{p=1}^{d} iH_p \varphi_p\right),
\label{10}
\end{equation}
where d is the number of parameters, while the following operator describes the sequential protocol
\begin{equation}
U = \prod_{p=1}^{d} \exp(iH_p \varphi_p).
\label{11}
\end{equation}
While these methods are generally different, they become equivalent when all operators are commutative (\([H_p, H_q] = 0\) for any \(p, q\)). In the following sections, we will focus on phase estimation in multimode optical interferometers, where \(H_p\) is a local operator acting on the p-th mode.\\
Thus, we consider that all the operators \(H_p\) commute, making it straightforward to demonstrate that \({H}_p =\mathcal{H}_p\). Therefore, the methods mentioned above are equivalent. To achieve the optimal estimation within the Cramér-Rao quantum limit, it is important to determine whether this lower bound can be reached. Generally, the quantum Cramér-Rao bound for multiparameter estimation can not be attained. However, in the case of pure states (\(\rho_{\text{in}} = |\psi_{\text{in}}\rangle\langle\psi_{\text{in}}|\)), it is possible to saturate the Cramér-Rao bound. In this context,  the condition for the Cramér-Rao bound is expressed as \cite{Matsumoto2002,Humphreys2013}
\begin{equation}
\text{Im}(\langle\psi_{\text{out}}|L_pL_q|\psi_{\text{out}}\rangle) = 0, ~~~~~\forall~ p, q
\label{12}
\end{equation}
here, \(|\psi_{\text{out}}\rangle = U|\psi_{\text{in}}\rangle\) represents the output state of the interferometer. A straightforward calculation shows that this condition is equivalent to
\begin{equation}
\langle\partial_p\psi_{\text{out}}|\partial_q\psi_{\text{out}}\rangle \in \mathbb{R}. ~~~~~~\forall~ p, q 
\label{13}
\end{equation}
When the operators \(H_p\) and \(H_q\) are commutative for all \(p, q\), it is evident that \cite{Liu2015,Liu2016}
\begin{equation}
\langle\partial_p\psi_{\text{out}}|\partial_q\psi_{\text{out}}\rangle = \langle\psi_{\text{in}}|H_pH_q|\psi_{\text{in}}\rangle = \langle\psi_{\text{in}}|H_qH_p|\psi_{\text{in}}\rangle, 
\label{14}
\end{equation}
and it can be deduced that the condition \eqref{14} is naturally satisfied. Therefore, the quantum Cramér-Rao bound is always achievable for the local parametrisation strategy.  In this context, the elements of the Fisher quantum information matrix, as described in equation \eqref {5}, can be expressed as follows 
\begin{equation}
F_{pq} = 4 (\langle\psi_{\text{in}}|H_pH_q|\psi_{\text{in}}\rangle - \langle\psi_{\text{in}}|H_p|\psi_{\text{in}}\rangle\langle\psi_{\text{in}}|H_q|\psi_{\text{in}}\rangle). \label{15}
\end{equation}
\section{Application to GHZ-type Photon-Added Coherent States}
In this section, we investigate independent phase estimation and the simultaneous estimation of multiple parameters using GHZ-type PACSs, and compare the resulting performances. In quantum physics, especially in quantum metrology, a Photon-Added Coherent State (PACS) plays a crucial role in enhancing the precision of estimating an unknown phase shift. This concept was introduced by Agarwal and Tara in 1991. The PACS acts as an intermediary state between the coherent state and the Fock state. It can be obtained by applying successively the creation operator $a^{+}$ m times on the Glauber coherent state $|{\alpha}\rangle$ \cite{Zavatta2004,Sivakumar1999,Jebli2022,Daoud2015,Kaydi2015}
\begin{equation}
|{\alpha}\rangle=\exp{\left(-\frac{1}{2}|\alpha|^{2}\right)}\sum_{k=0}^{+\infty}{\frac{{\alpha}^{k}}{\sqrt{k!}}|k\rangle}.
\label{16}
\end{equation}
In terms of coherent state $|{\alpha}\rangle$, the PACS $|{\alpha},n\rangle$ can be written as
\begin{equation}
|{\alpha},n\rangle=\frac{({a}^{+})^{n}}{\sqrt{\langle{\alpha}|({a}^{-})^{n}({a}^{+})^{n}|{\alpha}\rangle}}|{\alpha}\rangle,
\label{17}
\end{equation}
where 
\begin{equation}
{\langle{\alpha}|({a}^{-})^{n}({a}^{+})^{n}|{\alpha} \rangle}=n!L_{n}(-|\alpha|^{2}),
\label{18}
\end{equation}
with $L_{n}$ is the Laguerre polynomial of order m defined as follows
\begin{equation}
L_{n}=\sum_{k=0}^{n}{\frac{(-1)^{k}n!x^{k}}{(k!)^{2}(n-k)!}}.
\label{19}
\end{equation}
In terms of Fock states, the state $|{\alpha},n\rangle$ can be expressed as
\begin{equation}
|{\alpha},n\rangle=\frac{\exp(-|\alpha|^{2}/2)}{[n!L_{n}(-|\alpha|^{2})]^{\frac{1}{2}}}\sum_{k=0}^{\infty}\frac{\alpha^{k}\sqrt{(k+n)!}}{k!}|k+n\rangle,
\label{20}
\end{equation}
with a $\alpha=|\alpha|e^{i\phi}$ is a complex number.
PACSs are not orthogonal to each other because of
\begin{equation}
\langle{\beta}|(a^{-})^{n}(a^{+})^{n}|{\alpha}\rangle=e^{\frac{-|\alpha|^{2}}{2}}e^{\frac{-|\beta|^{2}}{2}}\sum_{i,j}^{\infty}\frac{(\alpha^{\ast}\beta)(i+n)!}{(i)^{2}}\neq{0}.
\label{21}
\end{equation}
For the value of $\beta=\alpha$, we find
\begin{equation}
\langle{\alpha}|(a^{-})^{n}(a^{+})^{n}|{\alpha}\rangle=n!L_{n}(-|\alpha|^{2}),
\label{22}
\end{equation}
and for $\beta=-\alpha$, the expression \eqref{21} becomes
\begin{equation}
\langle{-\alpha}|(a^{-})^{n}(a^{+})^{n}|{\alpha}\rangle=e^{-2|\alpha|^{2}}n!L_{n}(|\alpha|^{2}),
\label{23}
\end{equation}
The overlap of two PACSs $|{\alpha},n\rangle$ and $|-{\alpha},n\rangle$ is given by
\begin{equation}
\langle{-\alpha},n|{\alpha},n\rangle=\langle{\alpha},n|{-\alpha},n\rangle=e^{-2|\alpha|^{2}}\frac{L_{n}(|\alpha|^{2})}{L_{n}(-|\alpha|^{2})},
\label{24}
\end{equation}
this shows that the two-photon coherent states $|{\alpha},n\rangle$ and $|-{\alpha},n\rangle$, which have the same amplitude and a $\pi$ phase difference, are not orthogonal. Currently, we will now explore the quasi-GHZ coherent states, which are represented as
\begin{equation}
|GHZ_{l}(\alpha)\rangle=\mathcal{C}_{l}(\alpha)(|{\alpha}, {\alpha}, {\alpha}\rangle+e^{il\pi}|{-\alpha}, {-\alpha}, {-\alpha} \rangle),
\label{25}
\end{equation}
where $l\in \mathbb{Z}$, $l$=0 for a symmetric state, and l=1 for an antisymmetric state. the normalization constant $C_{l}(\alpha)$ is given by
\begin{equation}
\mathcal{C}_{l}(\alpha)={\bigr[{2+2{e^{-6|\alpha|^{2}}}\cos(l\pi)}\bigr]^{\frac{-1}{2}}}.
\label{26}
\end{equation}
Consequently, the photon-added quasi-GHZ coherent states are  rewritten as
\begin{equation}
    |GHZ_{l}(\alpha, n)\rangle= \mathcal{C}_{l}(\alpha, n)\Bigr[|{\alpha},n\rangle |{\alpha}\rangle |{\alpha}\rangle+e^{il\pi}|{-\alpha},n\rangle |{-\alpha}\rangle |{-\alpha}\rangle \Bigr],
    \label{27}
\end{equation}
where the normalization factor $\mathcal{C}_{l}(\alpha, n)$ reads
\begin{equation}
    \mathcal{C}_{l}(\alpha, n)=\Biggr[{2+2\cos(l\pi)e^{-6|\alpha|^{2}}\frac{L_{n}(|\alpha|^{2})}{L_{n}(-|\alpha|^{2})}}\Biggr]^{\frac{-1}{2}}.
    \label{28}
\end{equation}
When n=0, the state $|GHZ_{l}(\alpha, n)\rangle$ \eqref{27} reduces to $|GHZ_{l}(\alpha)\rangle$ \eqref{25}. Importantly, the overlap between the two states $|{\alpha}\rangle$ and $|-{\alpha}\rangle$ tends towards zero and becomes orthogonal when $|\alpha|$ goes to infinity. In this limiting scenario, the state simplifies to the standard GHZ-type three-qubit state.
\begin{equation}
   |GHZ_{l}(\infty)\rangle=\frac{1}{\sqrt{3}}[|{0\rangle}\otimes|{0\rangle}\otimes|{0\rangle}+e^{il\pi}|{1\rangle}\otimes|{1\rangle}\otimes|{1\rangle}],
    \label{29}
\end{equation}
where $|{0\rangle}\equiv|{\alpha\rangle}$ and $|{1\rangle}\equiv|{-\alpha\rangle}$.

\subsection{Independent estimation}
Entangled coherent states have been employed in various quantum metrology protocols, demonstrating their effectiveness in enhancing measurement accuracy beyond the standard quantum limit (SQL).  Among these states, photon-added quasi-GHZ coherent states have garnered considerable attention. In this work, we will focus specifically on local estimation,  particularly using the states of three qubits as probe states. We assume that the dynamics of the first qubit is governed by the local transformation \( e^{-i\varphi H} \equiv e^{-i\varphi H_1 \otimes I \otimes I} \), where \( H_1 \) is a local Hamiltonian governing the dynamics of qubit 1, and \( \mathbb{I} \) denotes the identity matrix. 
We start with independent estimation, which uses dedicated resources to estimate each parameter separately. This method is straightforward and serves as a basic comparison to evaluate the benefits of more advanced techniques, like simultaneous estimation. We will analyze this approach's accuracy using the Quantum Cramér-Rao Bound (QCRB) to understand the estimation limits.
Then, taking photon-added coherent states of GHZ-type \eqref{27} as input, the Quantum Fisher Information (QFI) for a parameter associated with the Hamiltonian \( H = a^{\dagger}a \) is given by
\begin{equation}
F=4[{\langle{\Psi_{\text{ind}}}|{H^{2}}|{\Psi_{\text{ind}}}\rangle-|\langle{\Psi_{\text{ind}}}|H|{\Psi_{\text{ind}}}\rangle|^{2}}],
    \label{31}
\end{equation}
with
\begin{equation}
    \begin{split}
     \langle{\Psi_{\text{ind}}}|{H^{2}}|{\Psi_{\text{ind}}}\rangle=\frac{2\mathcal{C}_{l}^{2}(\alpha, n)}{[n!L_{n}(-|\alpha|^{2})]}&
     \Bigl\{\Bigr[{(n+2)!L_{n+2}(-|\alpha|^{2})-3(n+1)!L_{n+1}(-|\alpha|^{2})+n!L_{n}(-|\alpha|^{2})}\Bigr]\\
     &+\cos(l\pi)e^{-6|\alpha|^{2}}\Bigr[{(n+2)!L_{n+2}(|\alpha|^{2})-3(n+1)!L_{n+1}(|\alpha|^{2})+n!L_{n}(|\alpha|^{2})}\Bigr]\Bigl\},
     \label{32}
    \end{split}
 \end{equation}
and
\begin{equation}
    \begin{split}
        \langle{\Psi_{\text{ind}}}|H|{\Psi_{\text{ind}}}\rangle=\frac{2\mathcal{C}_{l}^{2}(\alpha, n)}{[n!L_{n}(-|\alpha|^{2})]}&\Bigl\{\Bigr[{(n+1)!L_{n+1}(-|\alpha|^{2})-n!L_{n}(-|\alpha|^{2})}\Bigr]+\cos(l\pi)e^{-6|\alpha|^{2}}\\
        &\times\Bigr[{(n+1)!L_{n+1}(|\alpha|^{2})-n!L_{n}(|\alpha|^{2})}\Bigr]\Bigl\}.
        \label{33}
    \end{split}
\end{equation} 
Therefore, in the case of independent estimates with d parameters, the quantum Cramér–Rao bound of the state \eqref{27} is
\begin{equation}
|\delta\varphi|_{Ind}^{2}={d}{F^{-1}}.
\label{34}
\end{equation}
\subsection{Simultaneous estimation} 
After exploring independent estimation, where each parameter is treated separately, we turn to a more integrated approach: simultaneous estimation. This aims to infer all parameters in a single procedure, exploiting the quantum resources shared between modes. We assume the protocol evolves the system through the unitary transformation given in equation \eqref{10}. Then, we generalize the state in equation \eqref{27} to multi-mode photon-added GHZ-type coherent states for a multi-parameter estimation scenario. This extension is given by the following form 
\begin{equation}
    |\Psi_s\rangle=\mathcal{N}_{l}(\alpha, n, d)\Bigr[{|{\alpha},n\rangle_{0}\otimes^{d}_{i=1}|{\alpha}\rangle}_{i}+e^{il\pi}{|{-\alpha},n\rangle_{0}\otimes^{d}_{i=1}|{-\alpha}\rangle}_{i}\Bigr],
    \label{35}
\end{equation}
where 
\begin{equation}
    \mathcal{N}_{l}(\alpha, n, d)=\Biggr[{2+2\cos(l\pi)e^{-2(d+1) |\alpha|^{2}}\frac{L_{n}(|\alpha|^{2})}{L_{n}(-|\alpha|^{2})}}\Biggr]^{\frac{-1}{2}},
    \label{35}
\end{equation}
and $|{\alpha},n\rangle_{0}$ is the reference mode state (mode 0), which is a photon-added coherent state.
$\otimes^{d}_{i=1}{|{\alpha}\rangle}_{i}$ is the tensor product of coherent states $|\alpha\rangle$ in modes 1 to d.

In the context of the linear parameterization protocol, the Hamiltonian that governs the system's dynamics is given by the expression $H_i= a_{i}^{\dagger}a_{i}$. 
The terms in the QFI expression \eqref{15} for PACSs of type GHZ are expressed as
\begin{equation}
    \langle{\Psi_s}|{H_{p}H_{q}}|{\Psi_s}\rangle =\delta_{pq}b(\alpha, n, d)g(\alpha, n, d),\quad\quad  \langle{\Psi_s}|{H_{p}}|{\Psi_s}\rangle =b(\alpha, n, d)h(\alpha, n, d)
    \label{36}
\end{equation}
where the quantities $b(\alpha, n, d)$, \(g(\alpha, n, d)\) and \(h(\alpha, n, d)\) are given by the following expressions 

\begin{equation}
    \begin{split}
     b(\alpha, n, d)&=\frac{2\mathcal{N}_{l}^{2}(\alpha, n, d)}{[n!L_{n}(-|\alpha|^{2})]},\\
      g(\alpha, n, d)&=
     \Bigl\{\Bigr[{(n+2)!L_{n+2}(-|\alpha|^{2})-3(n+1)!L_{n+1}(-|\alpha|^{2})+n!L_{n}(-|\alpha|^{2})}\Bigr]\\
     &+\cos(l\pi)e^{-2d|\alpha|^{2}}\Bigr[{(n+2)!L_{n+2}(|\alpha|^{2})-3(n+1)!L_{n+1}(|\alpha|^{2})+n!L_{n}(|\alpha|^{2})}\Bigr]\Bigl\},\\
      h(\alpha, n, d)&=\Bigl\{\Bigr[{(n+1)!L_{n+1}(-|\alpha|^{2})-n!L_{n}(-|\alpha|^{2})}\Bigr]+\cos(l\pi)e^{-2d|\alpha|^{2}}\\
        &\times\Bigr[{(n+1)!L_{n+1}(|\alpha|^{2})-n!L_{n}(|\alpha|^{2})}\Bigr]\Bigl\}.
     \label{37}
    \end{split}
 \end{equation}

According to equation \eqref{15}, the elements of QFIM are defined as follows
\begin{equation}
    F_{pq} = 4\left[\delta_{pq} b(\alpha, n, d)g(\alpha, n, d) - b(\alpha, n, d)^2 h(\alpha, n, d)^{2}\right],
    \label{38}
\end{equation}
Therefore, the QFIM can be expressed as
\begin{equation}
    F=4b(\alpha, n, d){g}(\alpha, n, d) \biggl[{\mathbb{I}-\frac{b(\alpha, n, d){h}(\alpha, n, d)^{2}}{{g}(\alpha, n, d)}I}\biggl],
    \label{39}
\end{equation}
where  
$\mathbb{I}$ represents the identity matrix, and $I$ refers to a matrix in which every element is equal to 1, specifically $I_{pq}=1$ $(\forall{p, q})$.\\ 
By performing some calculations, the lowest QCRB is determined using the following total variance equation 
\begin{equation}
    |\delta\varphi|_{L}^{2}=\frac{d(\sqrt{d}+1)^{2}{h}(\alpha, n)^2}{4{g}(\alpha, n)^2}.
        \label{40}
\end{equation}
\paragraph*{}
Thus, for the non-linear parameterization protocol $H_i =(a_{i}^{\dagger}a_{i})^2$,  the QCRB is given by
\begin{equation}
    |\delta\varphi|_{NL}^{2}=\frac{d(\sqrt{d}+1)^{2}{s(\alpha, n, d)^{2}}}{4{r(\alpha, n, d)^{2}}},
     \label{41}
\end{equation} 
where the two quantities $r(\alpha, n, d)$ and $s(\alpha, n, d)$ are given by
\begin{align*}
r(\alpha, n, d)&=\left\lbrace \left[ (n+4)!L_{n+4}(-|\alpha|^{2})
	-10(n+3)!L_{n+3}(-|\alpha|^{2})+
	25(n+2)!L_{n+2}(-|\alpha|^{2})\right.\right.  \\
	&\left.\left. -15(n+1)!L_{n+1}(-|\alpha|^{2})+n!L_{n}(-|\alpha|^{2})\right]+\cos(l\pi)e^{-2d|\alpha|^{2}}\Bigr[(n+4)!L_{n+4}(|\alpha|^{2})\right. \\
	&\left. -10(n+3)!L_{n+3}(|\alpha|^{2})+
	25(n+2)!L_{n+2}(|\alpha|^{2})-15(n+1)!L_{n+1}(|\alpha|^{2})+n!L_{n}(|\alpha|^{2})\Bigr]\right\rbrace ,
\end{align*}
 \begin{align}
     s(\alpha, n, d)&=\Bigl\{\Bigr[{(n+2)!L_{n+2}(-|\alpha|^{2})-3(n+1)!L_{n+1}(-|\alpha|^{2})+n!L_{n}(-|\alpha|^{2})}\Bigr]\notag\\
    &+\cos(l\pi)e^{-2d|\alpha|^{2}}\Bigr[{(n+2)!L_{n+2}(|\alpha|^{2})-3(n+1)!L_{n+1}(|\alpha|^{2})+n!L_{n}(|\alpha|^{2})}\Bigr]\Bigl\}.
     \label{42}
 \end{align}

\subsection{Comparison of QCRB Limits for GHZ-Type PACS}
Here, we compare the quantum Cramer-Rao limits obtained through PACS of GHZ-Type in independent estimation, linear, and non-linear parameterization strategies. From expressions \eqref{40} and \eqref{41}, it is clear that all these limits are proportional to $d(\sqrt{d}+1)^{2}$. Additionally, the limits $|\delta\varphi|_{L}^{2}$, $|\delta\varphi|_{NL}^{2}$ have a correction of order O(d) compared to the independent estimate with PACS of GHZ-Type.\par 
The graphs illustrate the various values of the QCRB as a function of $d$ and $|\alpha|^{2}$ for both the symmetric case $(l=0)$ and the antisymmetric case $(l=1)$, considering different values of photon excitation number \(n\) $(n=0, 1, 4, 7, 10)$. The graphs were plotted for independent and simultaneous estimation. For the graphs plotted as a function of $|\alpha|^2$, we set the parameter d to the value 5. Conversely, for the graphs presented as a function of d, we set the amplitude \(|\alpha|^2\) to value 4.

\begin{figure}[h]
{{\begin{minipage}[b]{.24\linewidth}
\centering
\includegraphics[scale=0.321]{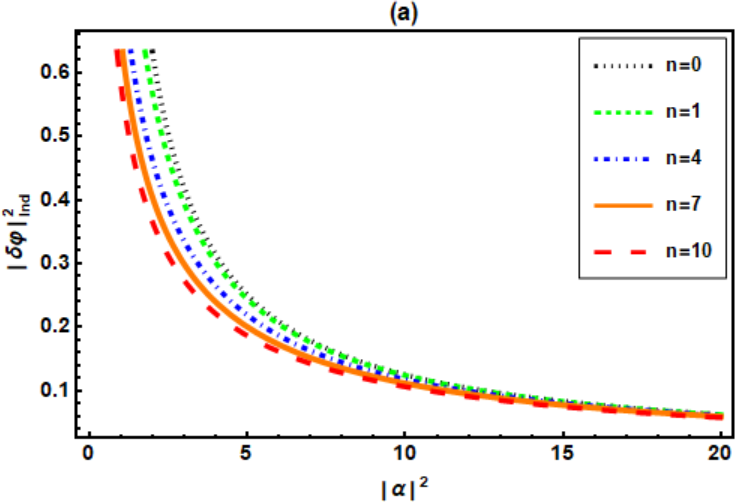} 
\end{minipage}\hfill
\begin{minipage}[b]{.25\linewidth}
\centering
\includegraphics[scale=0.325]{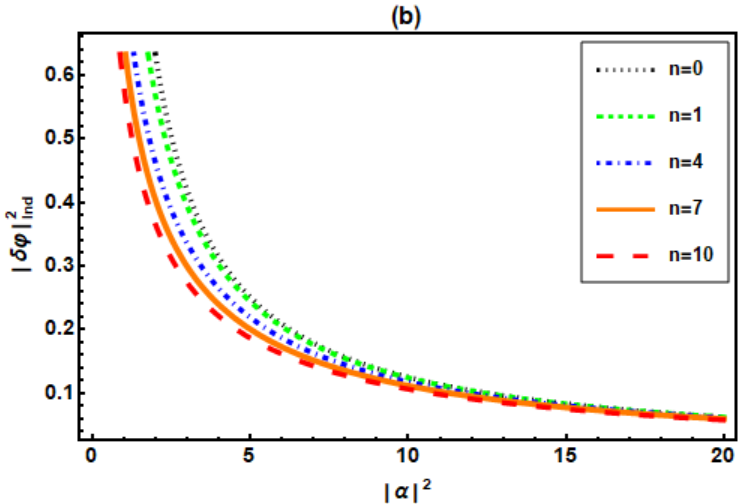}
\end{minipage}
\begin{minipage}[b]{.25\linewidth}
\centering
\includegraphics[scale=0.324]{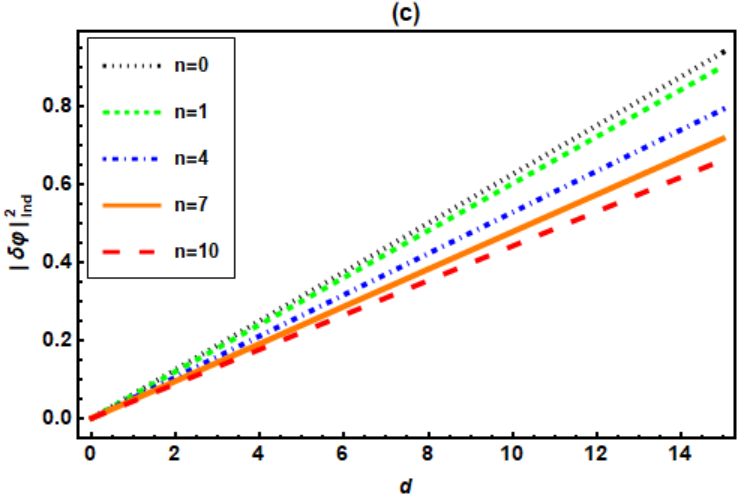}
\end{minipage}
\begin{minipage}[b]{.24\linewidth}
\centering
\includegraphics[scale=0.33]{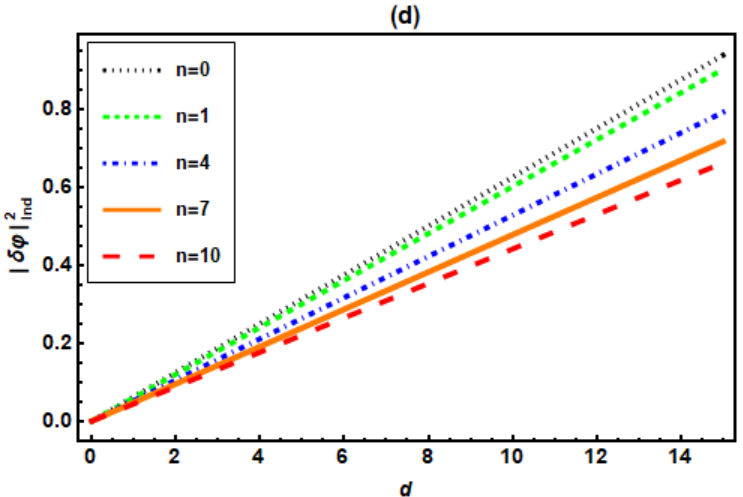}
\end{minipage}}}
  \caption{ The variation of ${|\delta\varphi|_{Ind}^{2}}$
versus the parameters ${|\alpha|^2}$ and {d} for  various values of photon excitation number {n}. \textbf{(a)} \textbf{d=5} and \textbf{l=0}, \textbf{(b)} \textbf{d=5} and \textbf{l=1}, \textbf{(c)} \textbf{$|\alpha|^{2}$=4} and \textbf{l=0}. \textbf{(b)} \textbf{$|\alpha|^{2}$=4} and \textbf{l=1}.}
	\label{fig1}
\end{figure}

Figures \ref{fig1}(a) and \ref{fig1}(b) depict the scenario of independent estimation that involves the amplitude of the coherent Glauber states $|\alpha|^{2}$. This applies to symmetric cases ($l=0$) and antisymmetric cases ($l=1$). These figures show that as $|\alpha|^{2}$ increases with the number of photons $n$, the total variance $|\delta\varphi|_{Ind}^{2}$ decreases exponentially and reaches a minimum. This means that the GHZ-type PACS ensures maximum precision in this case. 
Additionally, as the mean photon number $|\alpha|^{2}$ increases, the minimum bounds associated with $n=0, 1, 4, 7$, and $10$ tend to become equivalent and approach zero.  The figures show that the results for the two cases—symmetric and antisymmetric—are similar.

The plots presented in figures \ref{fig1}(c) and \ref{fig1}(d) clearly illustrate how $|\delta\varphi|_{Ind}^{2}$  varies with respect to the parameter \( d \) across different values of \( n\). 
As the parameter $d$ decreases and \(n\) increases, the QCRB also decreases, leading to greater precision.  In particular, for a fixed d, increasing $n$ drastically reduces $|\delta\varphi|_{Ind}^{2}$: for example, at $d=12$ and $n=10$ achieves 0.53 compared to 0.75 for $n=0$. In contrast, when \( d \) increases, the precision improves significantly for higher values of \( n \). Moreover, the behaviour of $|\delta\varphi|_{Ind}^{2}$ is similar in both symmetric and antisymmetric states.

\begin{figure}[h]
{{\begin{minipage}[b]{.24\linewidth}
\centering
\includegraphics[scale=0.325]{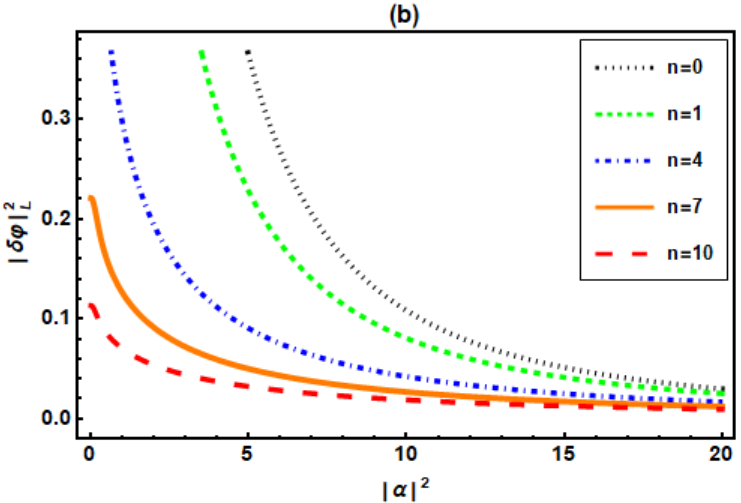} 
\end{minipage}\hfill
\begin{minipage}[b]{.25\linewidth}
\centering
\includegraphics[scale=0.325]{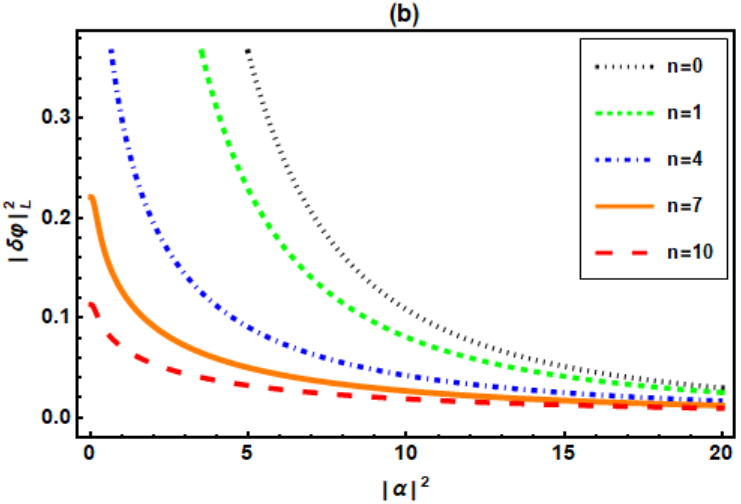}
\end{minipage}
\begin{minipage}[b]{.25\linewidth}
\centering
\includegraphics[scale=0.332]{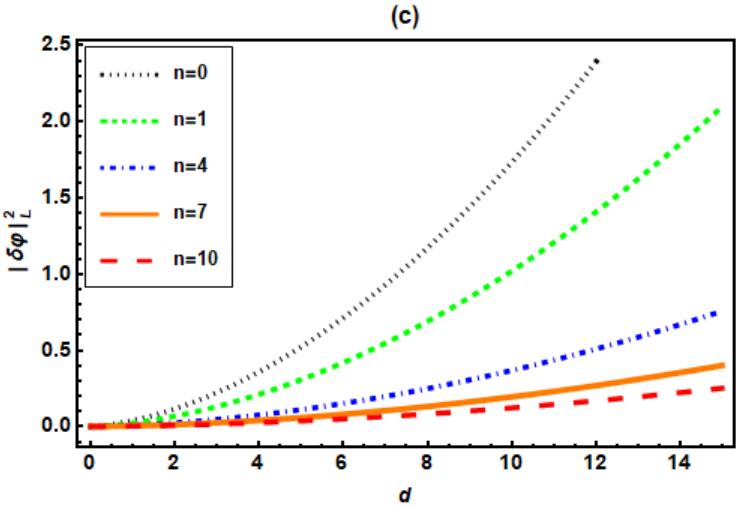}
\end{minipage}
\begin{minipage}[b]{.24\linewidth}
\centering
\includegraphics[scale=0.332]{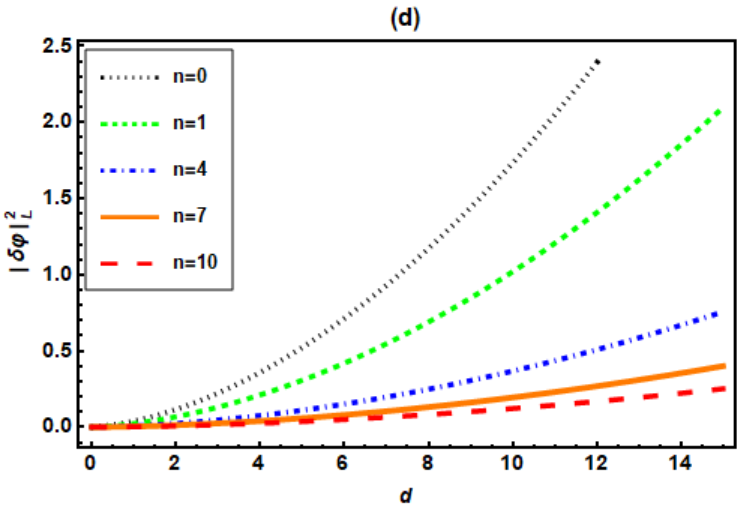}
\end{minipage}}}
  \caption{ The variation of $|\delta\varphi|_{L}^{2}$
		versus the parameters ${|\alpha|^2}$ and {d} for various values of photon excitation number {n}. \textbf{(a)} \textbf{d=5} and \textbf{l=0}, \textbf{(b)} \textbf{d=5} and \textbf{l=1}, \textbf{(c)} \textbf{$|\alpha|^{2}$=4} and \textbf{l=0}. \textbf{(b)} \textbf{$|\alpha|^{2}$=4} and \textbf{n=1}.}
	\label{fig2}
\end{figure}

For simultaneous estimation, particularly in the case of linear parameterization (as illustrated in Figures \ref{fig2}), the bound $|\delta\varphi|_{L}^{2}$ decreases exponentially as both the amplitude $|\alpha|^{2}$ and the parameter $n$ increase, indicating an improvement in precision (see Figures \ref{fig2} (a) and \ref{fig2} (b)). For example, at lower values of $n$ (e.g. \( n = 0, 1 \)), the bounds $|\delta\varphi|_{L}^{2}$ are $0.36$ and $0.22$, respectively, for $|\alpha|^{2}=5$. However, at higher \( n \) (e.g. \( n = 7, 10 \)), the value of \( |\delta\varphi|_{\text{Ind}}^{2} \) decreases significantly to approximately \(0.049\) and \(0.032\), respectively, under the same value of $|\alpha|^{2}$.

Furthermore, for a fixed value of \( |\alpha|^{2} \), it is evident from Figures \ref{fig2}(c) and \ref{fig2}(d) that the variance is minimized as \( n \) increases and \( d \) decreases. For example, when $n=7$, the value of $|\delta\varphi|_{L}^{2}$ is approximately $0.19$ for $d=10$, but this value decreases to $0.05$ as $d$ decreases to $5$.

The figures presented in this study, specifically Figures \ref{fig2}(a) and \ref{fig2}(b), indicate that the estimation accuracy is somewhat higher in the antisymmetric states compared to the symmetric states, particularly in the regions where \( n \) is large and \( |\alpha|^2 \) is small. However, the accuracy is nearly comparable for small values of $n$. For instance, at $n=7$ and $|\alpha|^{2}=1$, the Variance $|\delta\varphi_{L}|$ is approximately $0.125674$ for the symmetric state and $0.125677$ for the antisymmetric state, reflecting minimal differences in precision at this intermediate $n$. This suggests that the advantage of antisymmetric states becomes more pronounced only at significantly larger $n$. Conversely, the behaviour of the QCRB, as depicted in figure \ref{fig2}(c) and \ref{fig2}(d), is similar for both symmetrical and antisymmetrical states.

\begin{figure}[h]
{{\begin{minipage}[b]{.24\linewidth}
\centering
\includegraphics[scale=0.335]{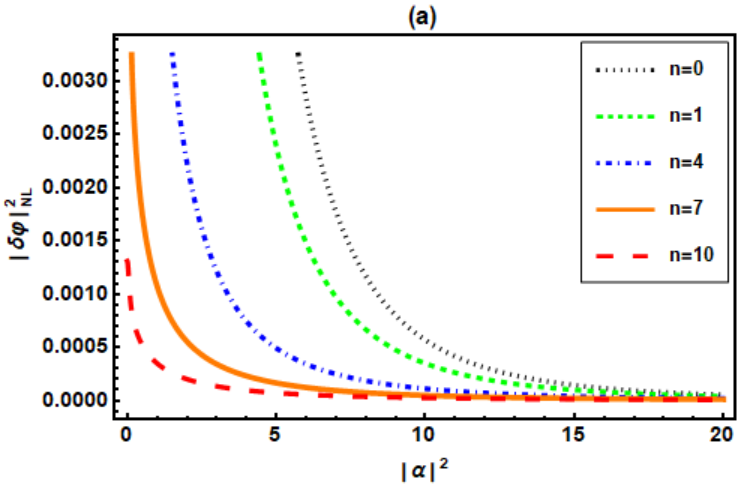} 
\end{minipage}\hfill
\begin{minipage}[b]{.25\linewidth}
\centering
\includegraphics[scale=0.337]{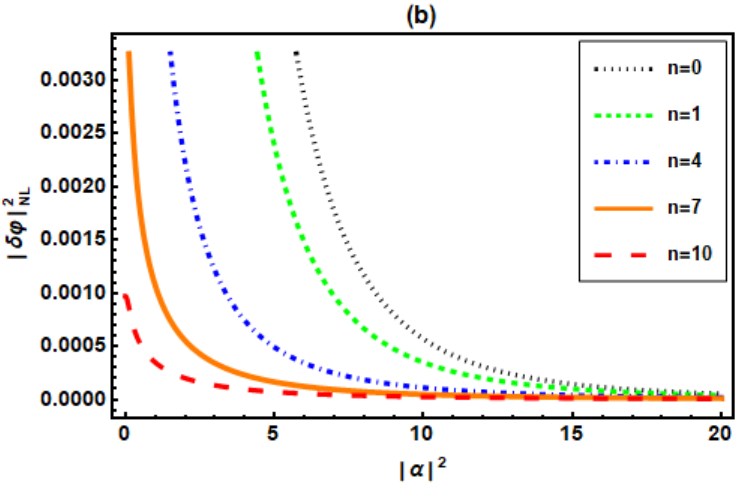}
\end{minipage}
\begin{minipage}[b]{.242\linewidth}
\centering
\includegraphics[scale=0.33]{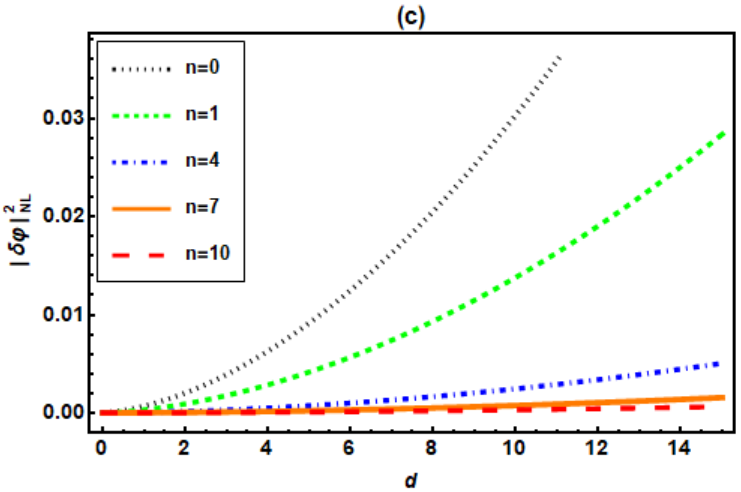}
\end{minipage}
\begin{minipage}[b]{.24\linewidth}
\centering
\includegraphics[scale=0.335]{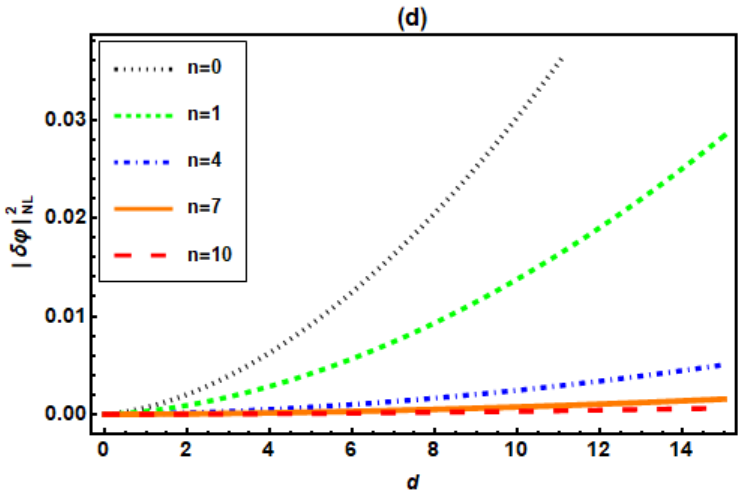}
\end{minipage}}}
  \caption{ The variation of $|\delta\varphi|_{NL}^{2}$
		versus the parameters ${|\alpha|^2}$ and {d} for various values of photon excitation number {n}. \textbf{(a)} \textbf{d=5} and \textbf{l=0}, \textbf{(b)} \textbf{d=5} and \textbf{l=1}, \textbf{(c)} \textbf{$|\alpha|^{2}$=4} and \textbf{n=0}. \textbf{(b)} \textbf{$|\alpha|^{2}$=4} and \textbf{l=1}.}
	\label{fig3}
\end{figure}

Finally, let's examine the scenario of non-linear parameterization. In this case, the behaviour of \( |\delta\varphi|_{NL}^{2} \) as a function of \( |\alpha|^{2} \) (shown in Figures \ref{fig3}(a) and \ref{fig3}(b)) and the parameter \(d\) (depicted in Figures \ref{fig3}(c) and \ref{fig3}(d)) closely resembles that of the linear parameterization case, although there is a difference in the scale of the variance. For figures \ref{fig3}(a) and \ref{fig3}(b), the minimum variance bound reaches a scale of \(10^{-3}\), while in figures \ref{fig3}(c) and \ref{fig3}(d), it reaches \(10^{-2}\).
These scales are significantly smaller than those in the linear parameterization case, demonstrating that the non-linear approach consistently outperforms the linear one for the same input state. For instance, at \(n=7\) and \(|\alpha|^2=1\) (as an example), the quantum Cramér-Rao bound is \(0.12\) for linear parameterization, whereas it drops to \(0.0011\) for non-linear parameterization.  

As a result, when comparing the limits \( |\delta\varphi|_{Ind}^{2} \), \( |\delta\varphi|_{L}^{2} \), and \( |\delta\varphi|_{NL}^{2} \), it is evident that simultaneous estimation consistently outperforms independent estimation for the same state. When considering the limits \( |\delta\varphi|_{L}^{2} \) and \( |\delta\varphi|_{NL}^{2} \) within simultaneous estimation, the photon-added coherent state (PACS) of GHZ-type demonstrates a lower limit with the nonlinear protocol.  

On the other hand, the PACS-type GHZ state surpasses the quantum standard limit, achieving Heisenberg limit precision \( \delta\varphi = \frac{1}{\bar{N}} \), where \( \bar{N} = \langle a^{\dagger}a \rangle \) represents the total average photon number.
This value changes with variations in the photon excitation number \(n\) and the amplitude \( |\alpha|^2 \). For large \( |\alpha|^2 \), the limits \( |\delta\varphi|_{L}^{2} \) and \( |\delta\varphi|_{NL}^{2} \) are approximately equivalent,  and the total average photon number is approximately $|\alpha|^2-1$.

\section{Comparison of ECS and NOON intricate states with PACS-type GHZ states}
\begin{figure}[h]
{{\begin{minipage}[b]{.48\linewidth}
\centering
\includegraphics[scale=0.37]{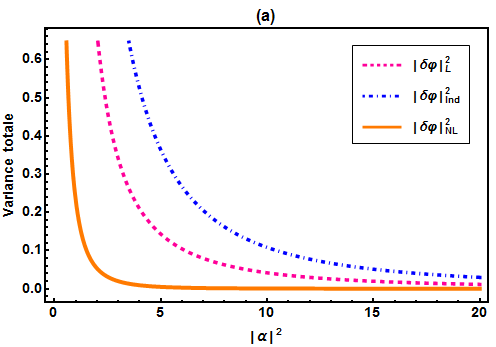} 
\end{minipage}\hfill
\begin{minipage}[b]{.48\linewidth}
\centering
\includegraphics[scale=0.37]{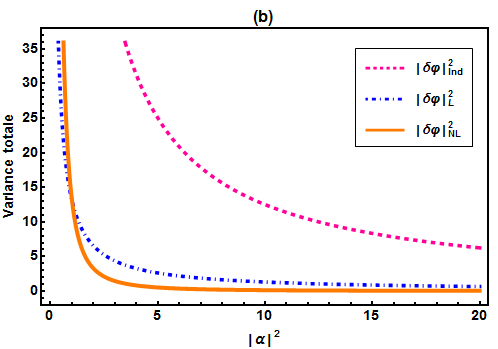} 
\end{minipage}}}
 \caption{The variation of $|\delta\varphi|_{Ind}^{2}$, $|\delta\varphi|_{L}^{2}$, and $|\delta\varphi|_{NL}^{2}$ as a function of the amplitude $|\alpha|^{2}$ with the total parameter number is set to  d = 5 here. a.) Represent the entangled coherent state ECS. b.) Represent the case of the NOON states.}
    \label{fig7}
\end{figure}

\begin{figure}[h]
{{\begin{minipage}[b]{.48\linewidth}
\centering
\includegraphics[scale=0.37]{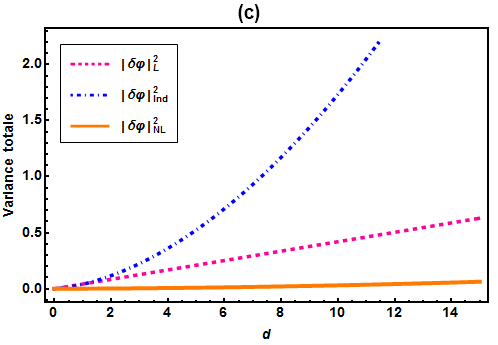} 
\end{minipage}\hfill
\begin{minipage}[b]{.48\linewidth}
\centering
\includegraphics[scale=0.37]{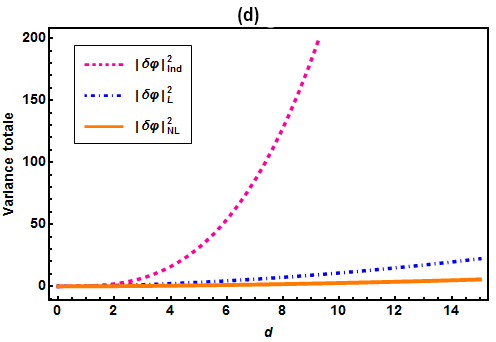} 
\end{minipage}}}
 \caption{The variation of $|\delta\varphi|_{Ind}^{2}$, $|\delta\varphi|_{L}^{2}$, and $|\delta\varphi|_{NL}^{2}$ as a function of the amplitude $d$ with the amplitude is set to $|\alpha|^{2}=4$ here. c.) Represent the case of an entangled coherent state ECS. d.) Represent the case of the NOON states.}
    \label{fig8}
\end{figure}

According to figures \ref{fig7} and \ref{fig8}, we have found that simultaneous estimation with the generalized entangled coherent state (ECS) can yield higher accuracy than independent estimation. Furthermore, we have observed that the entangled coherent states produce smaller bounds than the NOON states. Specifically, for a smaller number of photons, the generalized entangled coherent states for the linear protocol provide improved accuracy compared to the generalised NOON states in contrast to the non-linear protocol \cite{Humphreys2013,Liu2016}.

When comparing the latest results to previous ones, we observe that the PACS type GHZ provides the highest precision, especially as the number of photons \( n \) and \( |\alpha|^2 \) increase. For example, in the case of independent estimation, when \( n \) equals 4 and \( |\alpha|^2 \) is 4, the total average photon number is \( \Bar{N} = 10.38 \). In this scenario, the Heisenberg limit is approximately \( \delta \varphi_{\text{HL}} \simeq 0.09 \), while the standard quantum limit is about \( \delta\varphi_{\text{SQL}} \simeq 0.31 \). The Quantum Cramér-Rao Bound in this case is 0.22, which exceeds the standard quantum limit.

If we increase \( n \) to 10 and \( |\alpha|^2 \) to 10, the Heisenberg limit decreases to approximately 0.03. In this situation, the QCRB is 0.14, which surpasses the standard quantum limit: 0.19. This illustrates that the Heisenberg limit varies with changes in parameter values or estimation protocols, and these observations also apply to cases of simultaneous estimation.

\section{Homodyne detection}

Homodyne detection is an effective technique in both classical and quantum optics. It is known for its high sensitivity to phase measurements and ability to extract information encoded in the quantum states of light, particularly for measuring the quadratures of an electromagnetic field.
It works by mixing the signal to be measured with a reference wave, called a local oscillator, which has the same frequency as the signal. In this device, the signal is combined with the local oscillator using a semi-reflective beam splitter (50/50), thus generating two mixed beams. Each of these beams is then directed onto a photodiode, enabling differential detection. By adjusting the relative phase between the signal and the local oscillator, this set-up enables precise measurement of a specific quadrature of the electromagnetic field, thanks to the phase sensitivity of the detection system. This method is important for applications such as high-resolution imaging, quantum state characterization, and communication systems, due to its ability to detect small phase changes. 
In this section, we use homodyne detection as a measurement process. Through experimental homodyne detection, we perform quadrature measurements that provide a histogram of \( p_{\varphi} \) values. From this histogram, we derive a probability distribution known as the marginal distribution \( P(p|\varphi) \). This distribution characterizes the quantum fluctuations in the signal we aim to detect.\\

In the following, we consider the multi-mode photon-added GHZ-type coherent states \eqref{34}. We obtain
for the case of the linear protocol, the state $|\Psi_{out}\rangle$ of the system after the measurement as follows   
\begin{equation}
    |\Psi_{out}\rangle=\mathcal{N}_{l}(\alpha, n, d) e^{in\varphi}\Bigr[{|{\alpha}e^{i\varphi},n\rangle_{0}\otimes^{d}_{i=1}|{\alpha}\rangle}_{i}+e^{il\pi}|{-\alpha}e^{i\varphi},n\rangle_{0}\otimes^{d}_{i=1}{|{-\alpha}\rangle}_{i}\Bigr],
    \label{43}
\end{equation}
By performing some calculations, we have 
\begin{equation}
\langle{p|\alpha}\rangle=\Bigl(\frac{1}{\pi}\Bigl)^{\frac{1}{4}}\exp\left[ -\frac{p^2}{2}-i\sqrt{2}p\alpha+\frac{{\alpha}^2}{2}+\frac{|\alpha|^2}{2}\right] ,
    \label{44}
\end{equation} 
and 
\begin{equation}
\langle{p|\alpha},n\rangle=[n!L_{n}(-|\alpha|^{2})]^{-\frac{1}{2}}\left(\frac{i}{\sqrt{2}}\right)^{n}H_n\left(p+\frac{i\alpha}{\sqrt{2}}\right)\langle{p|\alpha}\rangle,
    \label{45}
\end{equation} 

In this section, we assume that $\alpha $ is a real  number and define $p = (p_0, p_1, ..., p_d)$ and $|p\rangle = |p_0, p_1, ..., p_d\rangle$, in which the indices represent the corresponding mode number. Indeed, the
output signal can be calculated by the mean value of the total quadrature operator
$\hat{p}_{tot} =\sum^{d}_{i=0}\hat{p}_i$. Consequently, the mean value of the total signal is 
\begin{equation}
\langle{\hat{p}_{tot}}\rangle=(1+d)\int_{-\infty}^\infty pP(p|\varphi) dp, 
    \label{46}
\end{equation} 
where $P(p|\varphi)=|\langle p, p, .., p|\Psi_{out}\rangle|^2$. The expression of $P(p|\varphi)$ to first order in ${\varphi}$ and after some simplification is given by 
\begin{equation}
    \begin{split}
           P(p|\varphi) &= {C}^2(\alpha, n, d) e^{-p^{2}(1+d)}\Biggl\{ \left[ e^{2\sqrt{2}p\alpha \varphi} H^2_n\left(p-\frac{\alpha}{\sqrt{2}}\right) + e^{-2\sqrt{2}p\alpha \varphi} H^2_n\left(p+\frac{\alpha}{\sqrt{2}}\right)\right]\\&+2cos(l\pi+2\sqrt{2}p\alpha(1+d))\left[ 
        H_n\left(p-\frac{\alpha}{\sqrt{2}}\right)H_n\left(p+\frac{\alpha}{\sqrt{2}}\right)\right]\\&-
        \sqrt{2}(n \alpha \varphi)  sin(l\pi+2\sqrt{2}p\alpha(1+d))\left[ 
        H'_{n}\left(p-\frac{\alpha}{\sqrt{2}}\right)H_n\left(p+\frac{\alpha}{\sqrt{2}}\right)+ H_n\left(p-\frac{\alpha}{\sqrt{2}}\right)H'_{n}\left(p+\frac{\alpha}{\sqrt{2}}\right)\right]\Biggl\},
    \end{split}
     \label{47}
\end{equation} 
where ${C}^2(\alpha, n, d)=\mathcal{N}_{l}^{2}(\alpha, n, d) \left(\frac{1}{\pi}\right)^{\frac{1+d}{2}}2^{-n} \left[n! L_n(-|\alpha|^2)\right]^{-1}e^{2\alpha^{2}(1+d)}$. The equation for error propagation is expressed by using the following formula
\begin{equation}   |\delta{\varphi}|^2=\frac{\langle{p_{tot}^{2}}\rangle-\langle{p_{tot}}\rangle^{2}}{|\partial_{\varphi}{\langle{p_{tot}}\rangle}|^2}.
    \label{48}
\end{equation}

The expression in equation \eqref{47} is quite complex, and calculating the expression of \eqref{46} is more difficult as it includes Hermite polynomials, exponential terms, and trigonometric terms. To simplify our analysis, we will treat this expression approximately and find reasonable formulas. For the case where \(\alpha\) is small, we find that
\begin{equation}
|\delta{\varphi}|^2_{hd}\quad{\sim}\quad\mathcal{N}^{-2}_{l}(\alpha, n, d)\left(\frac{1}{\pi}\right)^{-\frac{d}{2}}\frac{2n+1}{\alpha^{4}n^{2}}{(1+d)^{\frac{1}{2}}},
    \label{49}
\end{equation}
For the case where $\alpha$ is large, we can derive the following approximation 
\begin{equation}
|\delta{\varphi}|^2_{hd}\quad{\sim}\quad\mathcal{N}^{-2}_{l}(\alpha, n, d)\left(\frac{1}{\pi}\right)^{-\frac{d}{2}}\frac{1}{\alpha^{4n}n^{2}}{(1+d)^{\frac{3}{2}}} ,
    \label{50}
\end{equation}

From the variance expression, we can conclude that as $\alpha$ and $n$ increase, the variance decreases sharply, leading to improved accuracy. Conversely, when \(d\) increases, it rises significantly, leading to less accurate estimates. 
Thus, homodyne detection is sensitive to the d parameter compared to linear estimation and can be considered more powerful than the independent estimation strategy, in which the total variance is proportional to d.

Compared to homodyne detection, linear phase estimation with PACS of GHZ-type performs best in terms of the Cramér-Rao bound in the weak quantum regime (when $\alpha$ is minimal), or for multi-parameter systems. Homodyne detection remains a good compromise between performance and simplicity when working with intense coherent states (when $\alpha$ is maximal ).

\section{Conclusion}

In summary, we have studied multiparametric quantum metrology using photon-added multi-mode coherent states of GHZ-type. We provided explicit expressions for the quantum QCRB for independent, linear, and non-linear parameterization strategies. Our findings reveal that simultaneous estimation can achieve higher accuracy than independent estimation, particularly in the case of non-linear protocols. Furthermore, we analyzed the variation of the QCRB as a function of the amplitude $|\alpha|^{2}$ and the parameter $d$ for different photon excitation orders $n$ across all three protocols. The results show that precision improves as $|\alpha|^{2}$ increases and $d$ decreases. In both symmetric and antisymmetric cases, the variation in the QCRB versus the amplitude is identical for small $n$ values but diverges for larger $n$, with the antisymmetric case providing slightly higher precision. Similarly, when analyzed as a function of $d$, the variation remains comparable between the two states.\par

In addition, we compared PACS-based GHZ states with NOON states and entangled coherent states. Our results indicate that PACS-based GHZ states offer maximum precision, particularly as the photon excitation number $n$ increases. We conclude that independent and simultaneous estimation methods both have the potential to enhance quantum metrology, with each approach advantageous in specific scenarios. Independent estimation is particularly effective when parameters are weakly correlated, while simultaneous non-linear estimation is more suitable for systems with strongly correlated parameters and complex interdependencies.

Finally, we conclude that homodyne detection achieves performance close to quantum limits under certain conditions, but may be sub-optimal compared with strategies that fully exploit correlations between parameters, such as linear estimation. The choice of method will depend on available resources and accuracy objectives.

\end{document}